\documentstyle[11pt,newpasp_crete,twoside,epsf]{article}
\begin{document}
\title{Precession of Isolated Neutron Stars}
 \author{Bennett Link}
\affil{Department of Physics, Montana State University, Bozeman, MT
 59717, USA}

\begin{abstract}

I summarize the evidence for precession of isolated neutron stars and
theoretical effort to understand the observations. I discuss factors
that might set the precession period, describe constraints on the
material properties of the crust, and conclude with a brief discussion
of possible sources of stress that would deform a neutron star to the
extent required.

\end{abstract}

\section{Introduction}

Observations of precession of isolated neutron stars afford a variety
of interesting new probes of the physics of neutron stars. The
precession period is determined by the deformation of the star, and
hence by stresses in the crust and core. Observation of a clear
signature of damping of precession would constrain the dissipative
processes that enforce corotation between the crust and liquid
core. As a neutron star wobbles, the external torque that spins it
down could vary, giving a distinct timing signature. Precession could
therefore provide a direct probe, perhaps the {\em only} probe, of the
dependence of the spin-down torque on the angle between the magnetic
axis and the angular velocity. Also, as the neutron star wobbles, the
observer looks down the beam at different angles through the
precession cycle, allowing mapping of the beam morphology. A
rapidly-spinning precessing neutron star could be a strong source of
gravitational waves which might be detectable by LIGO or LISA.

\section{Observational Evidence for Precession}

The two isolated pulsars that provide the most compelling evidence for
precession are PSR 1642-03 (Cordes 1993; Shabanova, Lyne, \& Urama
2001) and PSR B1828-11 (Stairs, Lyne, \& Shemar 2000). PSR B1828-11 is
particularly convincing as it displays variations in pulse duration
and shape that are highly correlated with the timing residuals (see
Fig. 1), as one would expect for a wobbling neutron star. The pulse
width variations are about 5$^\circ$, suggesting a wobble angle (the
angle between the star's symmetry axis and the angular momentum) of
similar magnitude. The timing data are highly periodic, but
non-sinusoidal, indicating harmonic structure. Over 13 years, the
timing residuals show Fourier power at $\simeq 1000$ d, $\simeq 500$ d
and $\simeq 250$ d. The 1000-d component, however, does not appear
clearly in the Fourier analyses of the period, period derivative and
beam shape and so may not represent the precession period.

\begin{figure}

\plottwo{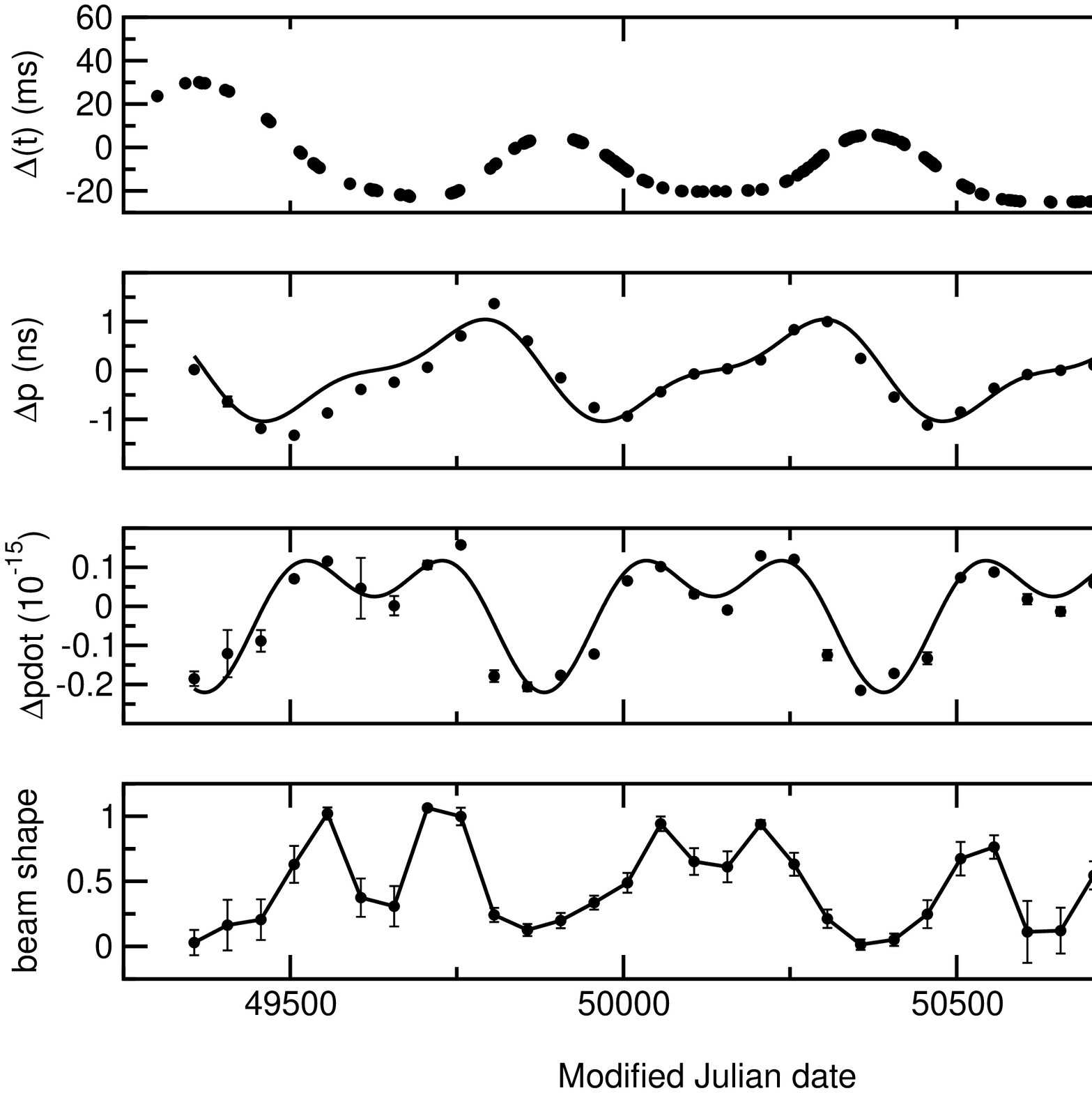}{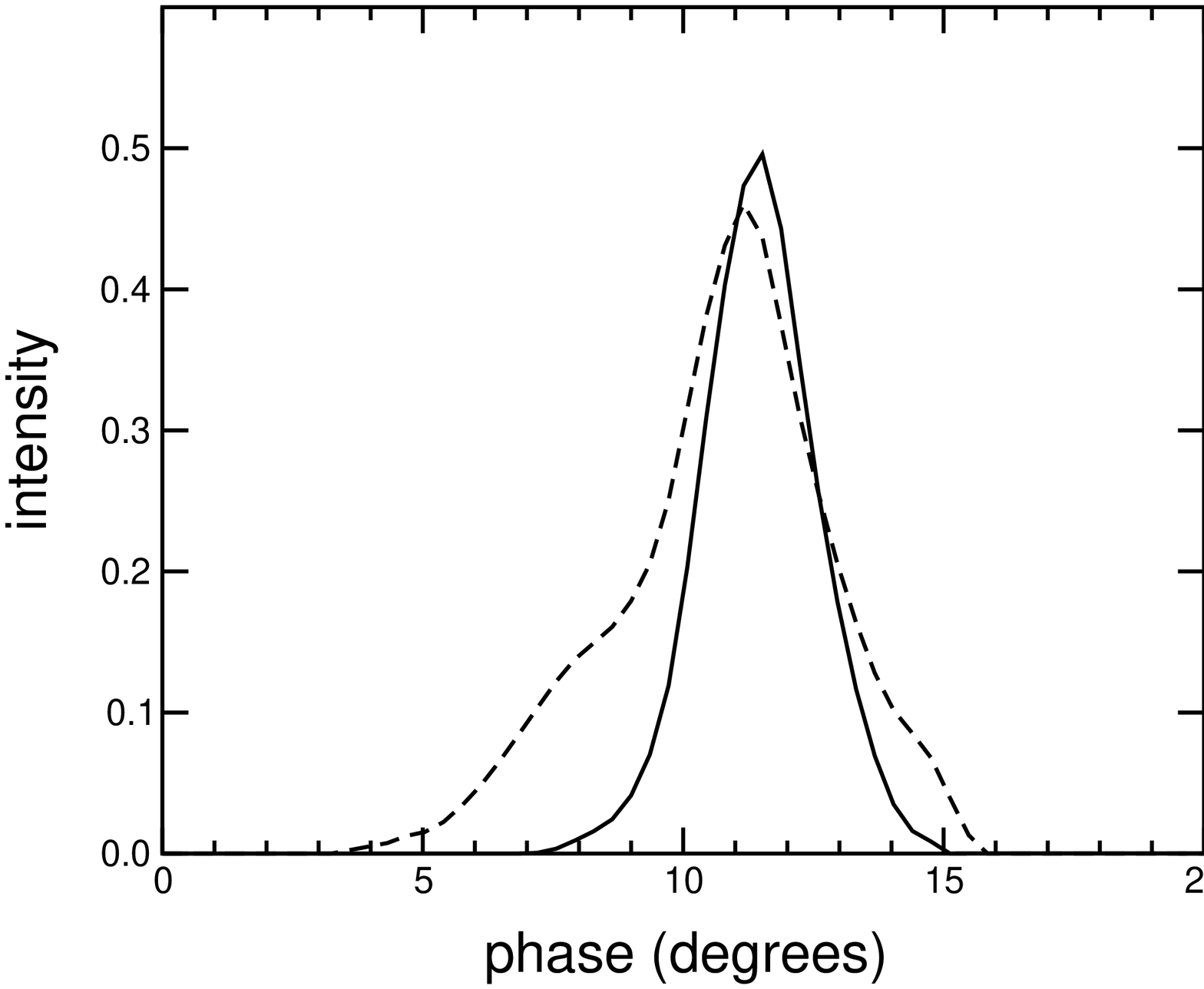}

\caption{The left panel shows the phase, period and period derivative
residuals for about 2000 days of data from PSR B1828-11. The data are
highly periodic but non-sinusoidal. The bottom graph shows the
correlated changes in beam width (the beam shape parameter is small
when the beam is wide, large when the beam is narrow). The solid
curves in the timing plots are fits from the model described in
the text. The right panel shows templates for the pulse profile in the wide
and narrow states. Data provided courtesy of I. Stairs.}

\end{figure}

\section{Theoretical Description of Precession}

\subsection{Free Precession}

Precession is the motion that results when a non-spherical rigid body
is set into rotation about any axis other than a principal axis.
Consider an oblate, biaxial, rigid object with principal moments of
inertia $I_3>I_2=I_1$. (The effects of triaxiality will be discussed
later). Let $I_3=I_1(1+\epsilon)$ where $\epsilon\ll 1$ is the
oblateness. Set the body rotating with angular velocity
$\mbox{\boldmath$\omega$}$ about an axis other than the symmetry axis
${\mathbf \hat s}$, as shown in Fig. 2. At any instant, the conserved
angular momentum ${\mathbf L}$, $\mbox{\boldmath$\omega$}$ and
${\mathbf\hat s}$ span a plane. The {\em wobble angle} $\theta_w$ and
$\vert\mbox{\boldmath$\omega$}\vert$ are both constants of the
motion. Free precession consists of a superposition of two rotations:
1) a fast wobble about ${\mathbf L}$ at approximately the spin rate,
and, 2) a slow, retrograde rotation about the symmetry axis at
frequency $\omega_p\simeq\epsilon\omega$ (for $\theta_w\ll 1$). With respect
to a coordinate system fixed in the body, $\mbox{\boldmath$\omega$}$
takes a circular path in a right-handed sense about $\mathbf{\hat s}$,
completing the circle in a precession period $p_p=2\pi/\omega_p$.

If a beam in direction ${\mathbf\hat{b}}$, taking an angle $\chi$ with
respect to ${\mathbf\hat{s}}$, is affixed to the body, a distant
observer will see a pulse when the beam passes through the plane
defined by the observer and ${\mathbf L}$. The observer will see
modulation of the pulse arrival times at frequency $\omega_p$,
accompanied by variations in pulse duration as she looks into the beam at
different angles. The phase variations have magnitude
$\delta\phi=\theta_w\cot\chi\sin\omega_pt$ (see, {\sl e.g.}, Nelson,
Finn, \& Wasserman 1990), corresponding to period variations of
\begin{displaymath}
{\delta p\over p} = {p\over p_p}\theta_w\cot\chi\cos\omega_pt, 
\end{displaymath}
where $p$ is the spin period. Assuming the precession period of PSR
B1828-11 is 500 d, and a wobble angle of 5$^\circ$, the observed
amplitude of the period variations (about 1 ns) implies $\chi\simeq
20^\circ$. These period variations, however, have no harmonics. I next
describe how the star's spin-down torque could introduce the 
harmonic structure observed in PSR B1828-11 and PSR 1642-03. 

\begin{figure}
\plotfiddle{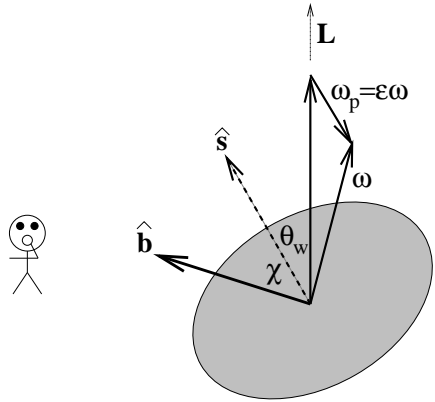}{5.cm}{0}{30}{30}{-100}{0}
\caption{The vectors describing precession of a freely-precessing,
biaxial object. The slow rotation about ${\mathbf\hat{s}}$ is
represented by the vector $\mbox{\boldmath$\omega_p$}$.} 
\end{figure}

\subsection{Precession Under Spin-down Torque}

The precession of an isolated neutron star is not truly torque-free; 
the star is being spun down by electromagnetic torque. If the torque
depends on the angle between the star's magnetic moment and the
angular velocity, then the spin-down torque will vary over the
precession period (Cordes 1993; Link \& Epstein 2001). The variable
torque will introduce variations in the star's spin rate (with respect
to the secular spin-down) which will add to the timing effects
described above. These variations are given by:
\begin{displaymath}
{1\over 2}I_1 {d\omega^2\over dt}\simeq \mbox{\boldmath$\omega$}\cdot{\mathbf
N}, 
\end{displaymath}
where ${\mathbf N}$ is the torque on the crust. Though the full
spin-down torque and its dependence on $\mbox{\boldmath$\omega$}$ are
unknown, let us consider the vacuum dipole torque (Davis \& Goldstein
1970) as an example:
\begin{displaymath}
{\mathbf N} = {2\omega^2\over 3 c^3}(\mbox{\boldmath$\omega$}\times {\mathbf
m})\times {\mathbf m}, 
\end{displaymath}
where ${\mathbf m}$ is the star's magnetic dipole moment.\footnote{The
near-field contribution to the torque, which does not affect the
star's spin rate, has been ignored.} The quantity
$\mbox{\boldmath$\omega$}\cdot{\mathbf N}$ has non-linear dependence
on the components of $\mbox{\boldmath$\omega$}$ which is particularly
strong for $\chi\simeq\pi/2$. Neglecting internal dissipation and
treating the precessing star as (effectively) a rigid body, this model
has three free parameters: the wobble angle $\theta_w$, the dipole
angle $\chi$ and the oblateness $\epsilon$. The solid curves shown 
in the timing plots of Fig. 1 are fits for $\theta_w=3.2^\circ$,
$\chi=89^\circ$ and $\epsilon=9.1\times 10^{-9}$ (Link \& Epstein
2001). The precession period is 511 d, with a harmonic at 256 d
arising from non-linearity in the torque. The derived wobble angle is
consistent with the observed pulse width variations. A simple model of
these variations allows crude mapping of the beam morphology, and
gives the hour-glass beam depicted in Fig. 3. This beam shape, though
non-standard, is similar to that found by Weisberg \& Taylor (2002)
for the binary pulsar B1913+16.

\begin{figure}
\plotfiddle{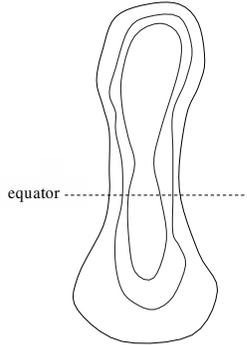}{4.cm}{0}{40}{40}{-150}{-50}
\caption{A schematic representation of the beam shape inferred for PSR
B1828-11.}
\end{figure}

\section{The Precession Period}

For a neutron star (or any object) to precess, it must have a
deformation axis which cannot follow the instantaneous spin vector. If
the moment of inertia corresponding to this deformation is $\Delta
I_d$, the precession period is
\begin{displaymath}
p_p = p \left ({I_c\over\Delta I_d}\right ), 
\end{displaymath}
where $I_c$ is the moment of inertia of the crust plus any component
of the star that corotates with the crust over timescales less than
$p$. If the precession period of PSR B1828-11 is $\simeq 500$ d, then
$\Delta I_d/I_c\simeq 10^{-8}$. This deformation is far less than the
rotational deformation, which follows the instantaneous spin axis and
therefore does not affect the precession period. What might sustain a
deformation of $\Delta I_d/I_c\simeq 10^{-8}$?

\subsection{Precession of an Elastic, Relaxed Body}

An elastic, relaxed (i.e., unstressed) body can precess, since
rigidity will prevent a portion of the spin-induced bulge from
following the instantaneous rotation vector once precession is
excited. Consider an unprecessing, spinning star with some
rigidity. If the star is relaxed, it will have an excess moment of
inertia $\Delta I_\omega$ about its rotation axis which is the same as
that for a self-gravitating fluid at the same spin rate. If the star
is carefully spun down to zero angular velocity without cracking or
otherwise relaxing, the star will not become spherical, but will
remain oblate by some amount $\Delta I_d$ under the stresses that have
developed from the spin down. The principal axis of the star is
aligned with the original rotation vector. Now suppose the star is
spun up to its original spin rate but about a different axis. The star
will precess, because the built-in deformation is not aligned with the
new spin axis. The star is also relaxed (except for the stresses
associated with the precession itself). The built-in deformation is
some fraction of the original spin deformation:
\begin{displaymath}
\Delta I_d = b \Delta I_\omega, 
\end{displaymath}
where $b$ is a {\em rigidity parameter} in the range $0\le b\le1$. For a
fluid $b=0$, while for an infinitely rigid solid, $b=1$. The
precession period of a relaxed, rigid object is then
\begin{displaymath}
p_p = {p\over b\epsilon_{rot}}, 
\end{displaymath}
where $\epsilon_{rot}$ is the rotational oblateness of a
self-gravitating fluid with spin period $p$. For the Earth, $b$ is
about 0.7; that is, the Earth behaves more like a solid than a
liquid. For the Earth, the above expression gives $p_p=440$ d, which
is the period of the famous Chandler Wobble.

Might the precession of PSR B1828-11 be similar to the Chandler
Wobble? A calculation of $b$ for a neutron star, using a realistic
model for the internal structure, gives $b\simeq 2\times 10^{-7}$ for
reasonable equations of state (Cutler, Ushomirsky, \& Link 2002). This
value is a factor of $\sim 30$ smaller than the previous estimate of
Baym \& Pines (1971). The smallness of $b$ is because the
gravitational energy density of a neutron star far exceeds the crust's
shear modulus; consequently, the crust behaves much more like a liquid
-- although a slightly ``rigid liquid'' -- than a solid. The implied
precession period of PSR B1828-11, {\em if its crust is relaxed}, is
about 100 years (Cutler et al. 2002). Hence, if crust
rigidity is responsible for the observed precession, the crust {\em
must be significantly strained}.

\subsection{The Precession Period of a Stressed Crust}

Strain in the crust will arise naturally as the star spins down. As
stress builds, it can be partially relieved by crustquakes or plastic
flow. Terrestrial solids fail ({\sl i.e.,} crack or flow) when the
strain reaches a critical value $\theta_c$, typically in the range
$10^{-5}<\theta_c<0.1$. At present, the critical strain for a given
solid cannot be calculated from first principles, and so must be
determined empirically.  The critical strain for the stellar crust can
be constrained by calculating the strain field of a
spinning-down neutron star and determining how much strain is needed
to sustain a given amount of deformation. To account for the
precession period of PSR B1828-11, the average critical strain of the
crust must satisfy (Cutler et al. 2002) 
\begin{displaymath}
\theta_c \ge 5\times 10^{-5} \left ({p_p\over 511 d}\right )^{-1}
\left ({I_c/I\over 0.01} \right ), 
\end{displaymath}
where $I$ is the total moment of inertia of the star and a fiducial
value of $I_c$ comparable to crust moment of inertia has been
chosen. Even if $I_c$ is comparable to $I$, the inferred lower limit
on $\theta_c$ is not unreasonable (by terrestrial standards),
suggesting that crust rigidity is sufficient to sustain the required
deformation.

\section{Effects of Vortex Pinning}

The lattice of the inner crust is expected to coexist with superfluid
neutrons. A rotating superfluid, such as liquid helium, is threaded by
quantized vortex lines. In a neutron star, an attractive interaction
between nuclei and vortices exists which might pin the vortices to the
lattice (Anderson \& Itoh 1975). As originally pointed out by Shaham
(1975), pinning is disastrous for long-period precession; the vortex
array acts as a gyroscope which drives the star to precess with a
period that is {\em at most} $\simeq 100$ spin periods if most of
the vortices in the crust are pinned. Under these circumstances,
damping of the precession is also very quick (Sedrakian, Wasserman, \&
Cordes 1999). However, if PSR B1828-11 is precessing with a wobble
angle of $\simeq 3^\circ$, pinning is most likely {\em unstable},
because the forces exerted on the pinned vortex lattice in a
precessing star are sufficient to cause global unpinning (Link \&
Cutler 2002).

\section{Effects of Triaxiality} 

So far I have focused on precession of a biaxial star. However, given
that the evolution and relaxation of the crust is probably quite
complex, and that magnetic stresses might also significantly deform
the star (Wasserman 2002), a neutron star is almost certainly a {\em
triaxial} object. For a triaxial, precessing object, the wobble angle
and $\vert\mbox{\boldmath$\omega$}\vert$ are not conserved
quantities. Some aspects of the precessional dynamics are then determined
by the dimensionless parameter
\begin{displaymath}
k^2 = {\epsilon^\prime\over\epsilon
(\epsilon-\epsilon^\prime)}\tan^2\theta_0, 
\end{displaymath}
where $\epsilon^\prime$ ($<\epsilon$) is defined by
$I_2=I_1(1+\epsilon^\prime)$ and $\theta_0$ is the minimum value of the
wobble angle. For $k$ of order unity, which is possible for
sufficiently large $\epsilon^\prime$, the arrival time residuals have
strong harmonics of $\omega_p$ even for {\em free}
precession. However, the angle between the beam and the observer at
the time of the pulse varies enormously during the precession cycle -- 
by 50$^\circ$ or more. Hence, $k$ of order unity cannot be tolerated
as an explanation of PSR B1828-11's timing behavior, as we would lose
sight of the beam.

For small $\theta_0$, $k$ can be $\ll 1$ even for extreme 
triaxiality ($\epsilon\simeq\epsilon^\prime$). In this limit, the
biaxial solution presented in Section 3.2 changes only
slightly, and so is robust for significant triaxiality. 

\section{Summary and Discussion}

The interpretation that PSR B1828-11 is a precessing neutron star is
supported by a simple model of torque-assisted precession with a
wobble angle of $\simeq 3^\circ$. The harmonic structure seen in the
timing data could be produced by a torque that depends on the angle
between the rotation vector and the magnetic moment, as in the vacuum
dipole model. These results are essentially unchanged if the star is
significantly triaxial. This model, if correct, provides evidence that
the spin-down torque {\em does actually depend on the relative
orientation of the spin and dipole axes}, as is usually assumed. The
vacuum dipole model for the torque (though alternatives should
certainly be considered) requires a rather extreme dipole angle of
$\chi\simeq 89^\circ$. Perhaps this conclusion is telling us something
about the preferred rotational state of a precessing star.

A precession period of $\simeq 500$ d in PSR B1828-11 cannot be
explained if the star is relaxed. Rather, the star must be under
stresses that sustain deformation. Crustal stresses alone could
account for the deformation, though Wasserman (2002) has proposed a
different model of precession in which both magnetic stresses and
crustal stresses deform the star.

\end{document}